\documentclass{article}
\usepackage{spconf,amsmath,graphicx,amsfonts}
\usepackage{booktabs,arydshln}
\usepackage{siunitx}
\usepackage{microtype}
\usepackage[colorlinks,linkcolor=black,urlcolor=black,citecolor=black]{hyperref}

\let\oldbibitem\bibitem
\renewcommand{\bibitem}[1]{\oldbibitem{#1}\ninept}

\title{FoleyGen: Visually-Guided Audio Generation}
\name{Xinhao Mei$^{1,2}$\sthanks{Work done during an internship at Meta.}, Varun Nagaraja$^1$, Gael Le Lan$^1$, Zhaoheng Ni$^1$, Ernie Chang$^1$, Yangyang Shi$^1$, Vikas Chandra$^1$}

\address{$^1$Meta AI, USA,\\
         $^2$University of Surrey, Guildford, UK
}
\begin{document}
%
\maketitle

\begin{abstract}
Recent advancements in audio generation have been spurred by the evolution of large-scale deep learning models and expansive datasets. However, the task of video-to-audio (V2A) generation continues to be a challenge, principally because of the intricate relationship between the high-dimensional visual and auditory data, and the challenges associated with temporal synchronization. In this study, we introduce \textbf{FoleyGen}, an open-domain V2A generation system built on a language modeling paradigm. FoleyGen leverages an off-the-shelf neural audio codec for bidirectional conversion between waveforms and discrete tokens. The generation of audio tokens is facilitated by a single Transformer model, which is conditioned on visual features extracted from a visual encoder. A prevalent problem in V2A generation is the misalignment of generated audio with the visible actions in the video. To address this, we explore three novel visual attention mechanisms. We further undertake an exhaustive evaluation of multiple visual encoders, each pretrained on either single-modal or multi-modal tasks. The experimental results on VGGSound dataset show that our proposed FoleyGen outperforms previous systems across all objective metrics and human evaluations.

\end{abstract}
\begin{keywords}
Sound generation, audio-visual learning, video-to-audio generation, multimodal learning
\end{keywords}
\section{Introduction}
\label{sec:intro}

Recent years have seen remarkable breakthroughs in audio generation, powered predominantly by the evolution of large-scale deep learning models and datasets. Despite great achievements in text-to-audio \cite{liu2023audioldm, kreuk2023audiogen} and text-to-music \cite{liu2023audioldm2, copet2023musicgen} generation, video-to-audio (V2A) generation lags behind, standing as a promising yet under-explored area due to its inherent challenges. Video-to-audio generation is the task of generating congruent soundscapes for a given visual signal, which requires parsing visual data, identifying sound-emitting objects, and then crafting corresponding sounds. V2A models are useful in various applications, such as generating sound for movies as a computational Foley artist, enhancing immersive experiences in virtual reality applications, and assisting visually impaired individuals for better spatial awareness.

\begin{figure}[t]
    \centering
    \includegraphics[width=\columnwidth]{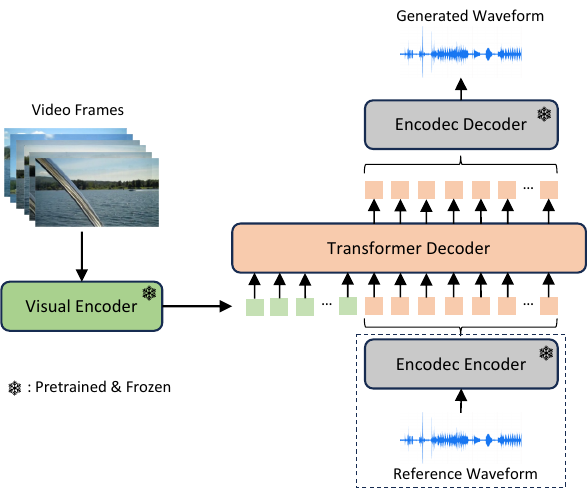}
     \caption{Overview of the FoleyGen system. The dashed-line block shows the EnCodec encoder for converting waveforms into discrete tokens, utilized only during training.}
    \label{fig:sys_overview}
\end{figure}

Achieving accurate and realistic V2A generation poses several challenges. First, the simultaneous interpretation of both visual and auditory data is intricate due to their respective high-dimensional natures. Second, real-world videos often contain visually irrelevant sounds where the objects emitting sound are absent from the visible frames. This discrepancy makes the generation of temporally synchronized audio extremely challenging. Finally, a single object can emit a diverse range of sounds depending on its interaction with varying environments, further complicating this task.

Initial efforts in V2A generation has predominantly focused on constrained visual contexts and a limited set of sound classes to simplify the problem \cite{Owens2016visually_indicated, chen2018VIG_dataset, zhou2018VEGAS_dataset}. Such approaches commonly utilized class-aware strategies \cite{chen2018VIG_dataset} or even trained separate models for distinct sound categories \cite{zhou2018VEGAS_dataset, chen2020regnet}. Consequently, these methods fail to generalize to open-domain videos. Recent advancements, however, indicate a rising interest in open-domain, visually guided audio generation. SpecVQGAN \cite{SpecVQGAN_Iashin_2021} and IM2WAV \cite{sheffer2023im2wav} both employ a language modeling method, leveraging the Transformer model to capture the joint distribution of visual features and discrete audio tokens encoded by vector-quantized variational autoencoder (VQ-VAE). In SpecVQGAN, the VQ-VAE operates specifically on spectrograms and subsequently employs a neural vocoder to convert generated spectrograms back into waveforms. In contrast, IM2WAV directly operates on waveforms, partitioning the VQ-VAE's latent space into two levels and utilizing dual Transformer models to model their respective distributions. Additionally, Diff-Foley \cite{luo2023diff_foley} introduces a latent diffusion method conditioned on contrastive audio-visual pretraining (CAVP) representations.

Inspired by the pioneering work of AudioGen \cite{kreuk2023audiogen} and MusicGen \cite{copet2023musicgen}, we introduce FoleyGen, a video-to-audio generation framework that adopts a language modeling paradigm. An overview of FoleyGen is provided in Figure~\ref{fig:sys_overview}. Specifically, our system encompasses three major components: a neural audio codec-EnCodec \cite{defossez2022encodec} for bidirectional conversion between audio and discrete tokens, a visual encoder for extracting visual features, and a Transformer model responsible for generating audio tokens conditioned on the visual context. Unlike SpecVQGAN \cite{SpecVQGAN_Iashin_2021}, the introduction of EnCodec provides better reconstruction quality and alleviates fidelity loss that often occurs during the spectrogram-to-waveform conversion process. Additionally, it eliminates the need for deploying multiple Transformer models IM2WAV \cite{sheffer2023im2wav}. A prevalent problem in V2A generation is the misalignment of generated audio with the visible actions in the video. To enhance the temporal alignment between visible actions and corresponding audio events, we propose and explore three different visual attention mechanisms. Furthermore, we conduct an exhaustive evaluation of various visual encoders, pretrained on both single-modal and multi-modal tasks. The experimental results show that our proposed FoleyGen outperforms previous systems across all objective metrics and human evaluations.

\section{Proposed Method}
\label{sec:method}
Given a video clip, a video-to-audio generation system is designed to produce an audio clip that is both semantically consistent with and temporally aligned to the accompanying video content. 
The video-to-audio generation process can be formulated as $\mathcal{H}: v\mapsto a$, where $v$ refers to the frames of a video input and $a$ corresponds to the generated audio waveform. Figure 1 presents the architecture of FoleyGen, our proposed system. FoleyGen comprises three main components: a neural audio codec for the bidirectional conversion between waveforms and discrete tokens, a visual encoder for feature extraction from video frames, and an audio language decoder tasked with generating discrete audio tokens based on the extracted visual features. This section first provides a detailed introduction to each major component of FoleyGen. To improve the temporal alignment of the visual input and generated audio, we propose using different visual attention mechanisms, which are described at the end of this section.

\subsection{Neural Audio Codec}
Modeling the distribution of time-domain waveforms presents significant challenges and computational inefficiencies, primarily due to their high-dimensional and lengthy characteristics. In audio generation systems, autoencoders are commonly utilized to encode audio waveforms into a latent space, which can be either continuous \cite{liu2023audioldm} or discrete \cite{kreuk2023audiogen}. Inspired by AudioLM \cite{borsos2023audiolm} and AudioGen \cite{kreuk2023audiogen}, we adopt EnCodec, a state-of-the-art neural audio codec \cite{defossez2022encodec}, for our experiments. EnCodec comprises an encoder that compresses audio waveforms into latent vectors, a residual vector quantizer (RVQ) for converting these latent vectors into discrete tokens, and a symmetric decoder that reconverts these tokens back into audio waveforms. Given an audio clip $\mathbf{a} \in \mathbb R ^{t \times f_s}$, where $t$ is the duration and $f_s$ is the sampling rate, the encoder first compresses $\mathbf{a}$ into a latent representation $\mathbf{z} \in \mathbb R^{L\times d}$. Here, $d$ is the dimensionality of the latent vector, and $L$ is the number of down-sampled time steps. A RVQ with $N_q$ codebooks then transforms the encoded latent vectors into $N_q \times L$ discrete tokens. The discrete audio tokens are further used as the representation of audio in the language modeling stage. The EnCodec decoder converts the generated audio tokens to waveforms. The EnCodec encoder is used only during training. We adhere to the same hyperparameter settings as outlined in the EnCodec paper, please refer to \cite{defossez2022encodec} for details. 

The adoption of EnCodec offers a high compression rate while keeping high reconstruction quality. Unlike other autoencoders that operate on spectrograms \cite{SpecVQGAN_Iashin_2021, luo2023diff_foley}, EnCodec eliminates the need for an additional vocoder and thus obviates the potential fidelity loss that may occur when converting a generated spectrogram back to a waveform.

\subsection{Visual Encoder}
Given a visual input $\mathbf{v} \in \mathbb R^{T \times C \times H \times W}$, where $T$ represents the number of frames (which can be 1 for a single image), $C$ is the number of channels, and $H$ and $W$ denote the height and width of the visual input, respectively, the visual encoder generates feature vectors $F\in \mathbb R^{T \times D}$ with $D$ being the number of dimension of the language decoder. The quality of the extracted visual features $F$ is critical for achieving semantically consistent and temporally aligned audio generation. A suboptimal visual encoder may lead to loss of important visual cues, resulting in an audio output that lacks fidelity or congruency with the original video content. To explore the efficacy of different visual encoders, we conducted a series of experiments using a variety of popular visual encoders trained with uni-modal and multi-modal tasks. These visual encoders include ViT \cite{dosovitskiy2021vit}, CLIP \cite{radford2021clip}, ImageBind \cite{girdhar2023imagebind} and VideoMAE \cite{tong2022videomae}.

\subsection{Audio Language Decoder}
Audio is represented as discrete tokens after being encoded by EnCodec \cite{defossez2022encodec}, therefore, the video-to-audio generation problem can be formulated as a conditional language modeling task. Given visual features extracted as conditional information, we employ a Transformer model \cite{vaswani2017transformer} to generate discrete audio tokens autoregressively. The Transformer model is decoder-only and omits the cross-attention block. The visual features are prepended to the sequence of audio tokens for conditioning. Due to EnCodec's residual vector quantization, each timestep encodes multi-stream tokens using residual codebooks. To effectively capture these multi-stream tokens, we adopt the delay pattern introduced in MusicGen \cite{copet2023musicgen}. This approach parallelly models multiple streams of audio tokens while maintains offsets between the streams. The incorporation of the delay pattern ensures high efficiency and eliminates the need for predicting tokens in a flattened pattern. Moreover, it sidesteps the requirement of multiple Transformer models \cite{borsos2023audiolm, sheffer2023im2wav}.

\begin{figure}[t]
    \centering
    \includegraphics[width=\columnwidth]{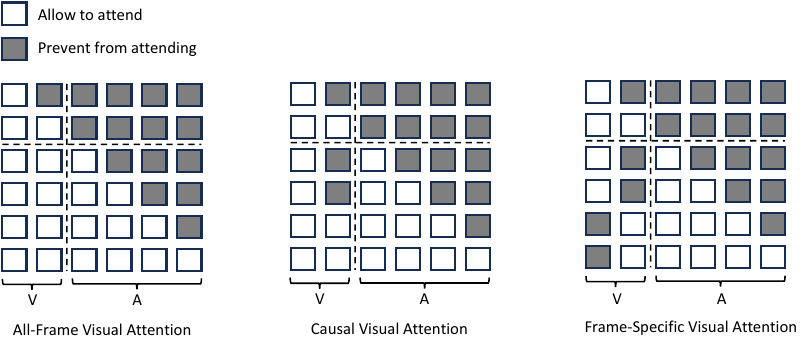}
     \caption{Overview of the three visual attention mechanisms. For simplicity, here we assume we have \num{2} visual features `V' and \num{4} audio tokens `A' with a frame rate of \num{2} Hz.}
    \label{fig:attention}
\end{figure}

\subsection{Visual Attention Mechanism}
Generating audio that is temporally aligned with a video presents significant challenges. To address this, we introduce and explore three distinct visual attention mechanisms. Figure~\ref{fig:attention} shows the overview of the three attention mechanisms.

\textbf{All-Frame Visual Attention}: In our baseline setting, we employ the default causal attention mechanism inherent in the Transformer decoder. Given that the visual features are prepended to the discrete tokens, during the generation process, the audio tokens have the capability to attend to all visual features. While this provides a broad context, it might confuse the model regarding the exact timing for sound generation due to an overabundance of visual information.

\textbf{Causal Visual Attention}: As a countermeasure, we investigate a ``causal" approach wherein, during the audio token generation, the model is restricted to attending only to visual frames that precede and align with the current timestep. This sequential attention might help the model to better synchronize the audio with the visual cues.

\textbf{Frame-Specific Visual Attention}: In a more restrictive approach, we introduce``frame-specific visual attention", where the model's attention is confined strictly to visual features of the concurrent time frame during generation. This strict attention mechanism ensures that the model generates audio only based on the current visual context.

\begin{table*}[!t]
\begin{center}
\caption{Experimental results on VGGSound dataset. Here we use all-frame visual attention.}
\label{table:tab_main_results} 
{\small
\begin{tabular}[\linewidth]{c | c |c c c | c c} 
 \toprule
 Methods & Visual Encoder & FAD $\downarrow$ & KL $\downarrow$ & IB (\%) $\uparrow$ & OVR (\%) $\uparrow$ & REL (\%) $\uparrow$ \\
 \midrule
 SpecVQGAN \cite{SpecVQGAN_Iashin_2021} & ResNet-50 & 6.64 & 3.10 & - & 5.6 & 5.6 \\
 IM2WAV \cite{sheffer2023im2wav} & CLIP & 6.41 & 2.54 & - & 16.7 & 31.1 \\
 \midrule
 Ours & CLIP & \textbf{1.65} & \textbf{2.35} & \textbf{26.1} & \textbf{77.7} & \textbf{63.3}\\
 
\bottomrule
\end{tabular}
}
\end{center}
\end{table*}

\begin{table}[!t]
\begin{center}
\caption{Experimental results on VGGSound dataset with models trained using different visual encoders.}
\label{table:tab_visual} 
{\small
\begin{tabular}[\columnwidth]{c | c c c} 
 \toprule
 Visual Encoder & FAD $\downarrow$ & KL $\downarrow$ & IB(\%) $\uparrow$ \\
 \midrule
 CLIP & \textbf{1.65} & 2.35 & 26.1\\
 ViT & 1.75 & 2.50 & 23.7\\
 ImageBind & 1.66 & \textbf{2.34} & \textbf{26.3}\\
 VideoMAE & 2.59 & 3.25 & 17.4\\
\bottomrule
\end{tabular}
}
\end{center}
\end{table}

\begin{table}[!t]
\begin{center}
\caption{Experimental results on VGGSound dataset with models trained using different attention mechanisms. The visual encoder used is CLIP.}
\label{table:tab_attention} 
\resizebox{\columnwidth}{!}{
\begin{tabular}[\columnwidth]{c | c c c | c c} 
 \toprule
 Attention & FAD $\downarrow$ & KL $\downarrow$ & IB(\%) $\uparrow$ & OVR (\%) $\uparrow$ & ALI (\%) $\uparrow$ \\
 \midrule
 All-frame & \textbf{1.65} & \textbf{2.35} & \textbf{26.1} & \textbf{63.3} & \textbf{55.6} \\
 Causal & 2.18 & 2.44 & 25.5 & 14.4 & 13.3 \\
 Frame-specific & 2.49 & 2.46 & 24.2 & 22.3 & 31.1 \\
\bottomrule
\end{tabular}
}
\end{center}
\end{table}

\section{Experiments}
\label{sec:exp}

\subsection{Dataset}
We target at open-domain visually guided audio generation. Therefore, we use the VGGSound \cite{chen2020vggsound} dataset, which contains around \num{200}k \num{10}-second video clips sourced from YouTube with diverse contents. Since some video clips are not downloadable anymore, our version contains \num{159318} samples in the train set and \num{13161} samples in the test set. 

\subsection{Implementation Details}
All the audio clips in the dataset are sampled to \num{16}k Hz monophonic audio. For the EnCodec, we follow the same downsampling strides $[2, 4, 5, 8]$ in the encoder, which leads to a frame rate of \num{50} Hz. We employ four codebooks with a codebook size of \num{2048}. For video data, we sample one frame per second and follow the prepocessing protocols (e.g., resize, normalize) in the visual encoders. A linear layer is used after the visual encoder to project the visual features to the same dimension of the Transformer model. The Transformer decoder consists of \num{24} layers with \num{16} heads and a dimension of \num{1024}. A memory efficient flash attention \cite{dao2022flashattention} is used to improve the speed and memory usage.

The models are trained for \num{20}k steps with a batch size of 256. AdamW optimizer with $\beta_1=0.9$, $\beta_2=0.95$, and a weight decay of \num{0.1} is used. The learning rate is set to \num{1e-4} and warm up is used in the first \num{4}k steps. In addition, classifier-free guidance \cite{ho2022classifier} is also employed to achieve better visual adherence. During training, the visual condition is dropped (i.e., replaced with null vectors) with a probability of \num{0.1}. During inference, the classifier-free guidance scale of \num{3.0} is used, and we employ top-k sampling with k setting to \num{256}. 

\subsection{Evaluation Metrics}
To evaluate the performance of FoleyGen, we carry out both objective and subjective evaluations. For objective evaluation, we employ Fr\'echet Audio Distance (FAD) \cite{kilgour2018fad}, Kullback-Leibler Divergence (KLD), and ImageBind (IB) score \cite{girdhar2023imagebind}. FAD calculates the distribution distance between the features of generated and reference audio clips, where the features are calculated using VGGish network \cite{hershey2017vggish} trained on AudioSet. KLD compares the label distribution of target and generated audio calculated by a pretrained PaSST model \cite{koutini22passt}. FAD demonstrates a strong correlation with human perception regarding audio quality, whereas KLD primarily captures the audio concepts present in the recording \cite{kreuk2023audiogen}. To evaluate the relevance between the generated audio and video, we propose using the ImageBind model \cite{girdhar2023imagebind} to compute a relevance score. Since ImageBind is trained to learn a joint embedding across six distinct modalities, the cosine similarity of its embeddings for both video and generated audio can capture semantic relevance between them. For subjective evaluation, human listeners are asked to compare samples generated by distinct models and identify the one that demonstrated superior performance based on specific criteria, which included overall quality (OVR), relevance (REL) to the corresponding visual input. Temporal alignment (ALI) is considered when evaluating the attention mechanisms.

\subsection{Results}
Table~\ref{table:tab_main_results} presents the primary results of our study, where we benchmark our proposed FoleyGen system against two previous state-of-the-art methods, SpecVQGAN \cite{SpecVQGAN_Iashin_2021} and IM2WAV \cite{sheffer2023im2wav}. Given that IM2WAV utilized FAD and KLD as evaluation metrics, we adopted their scores directly. For subjective evaluation, we generated samples using their pretrained models. It's evident from the results that FoleyGen consistently surpasses both SpecVQGAN and IM2WAV in both objective and subjective metrics. Notably, there's a marked reduction in the FAD score. The trends in subjective evaluations are congruent with the objective metrics. Several factors can be attributed to this improvement. First, the integration of EnCodec facilitates a heightened compression ratio of audio tokens and leads to a enhanced reconstruction quality. This elevated compression ratio simplifies the modeling of its distribution for the language model. Second, the utilization of the delay pattern in token generation avoids the need for multiple Transformer models, culminating in superior performance.

Table~\ref{table:tab_visual} shows the results of our models when trained using various visual encoders. It can be observed that visual encoders that are pre-trained via multi-modal tasks, (i.e., CLIP \cite{radford2021clip} and ImageBind \cite{girdhar2023imagebind}), exhibit comparable performances and surpass those trained solely on uni-modal tasks. ViT, which has been pre-trained through a discriminative task, outperforms VideoMAE. Since VideoMAE is trained using masked autoencoder with self-supervised learning, fine-tuning might be required when adopt it for downstream tasks.

Table~\ref{table:tab_attention} presents the results achieved using different attention mechanisms. All-frame visual attention notably surpassed the other two, both in objective metrics and human evaluations. Interestingly, while the frame-specific attention lagged in objective evaluations, it demonstrated an enhanced performance in human evaluation as compared with causal visual attention. However, a critical insight from human evaluations reveals that the systems still struggle with temporal alignment, and sometimes fail to capture prominent actions within the video.

\section{Conclusions}
\label{sec:conclusions}

In this paper, we introduced FoleyGen, a video-to-audio generation model following a language modeling paradigm. FoleyGen utilizes the EnCodec for bidirectional waveform-token conversion,a visual encoder for visual feature extraction and a Transformer decoder for conditioned audio token generation. Our evaluations demonstrate that FoleyGen surpasses prior methodologies in both objective metrics and human evaluations. Through our explorations, we observed that visual encoders trained on multimodal tasks exhibit superior performance. While we introduced visual attention mechanisms to enhance audio-video temporal alignment, it remains a persistent challenge in the domain. Future research should delve deeper into improving the temporal cohesion of video-to-audio generation systems.

\newpage
\bibliographystyle{IEEEbib}
\bibliography{refs}
\end{document}